\documentclass[10pt,conference]{IEEEtran}

\usepackage{algorithm}
\usepackage{algpseudocode}
\usepackage{cite}
\usepackage{amsthm}
\usepackage{amsmath,amssymb,amsfonts}
\usepackage{graphicx}
\usepackage{textcomp}
\usepackage{xcolor}
\usepackage{bm}
\usepackage{todonotes}
\usepackage{tabu}
\usepackage{balance}
\usepackage{comment}
\def\BibTeX{{\rm B\kern-.05em{\sc i\kern-.025em b}\kern-.08em
    T\kern-.1667em\lower.7ex\hbox{E}\kern-.125emX}}
\usepackage{acro}
\usepackage{mathtools}
\mathtoolsset{showonlyrefs}

\DeclareAcronym{3GPP}{short = 3GPP ,long = 3rd Generation Partnership Project}
\DeclareAcronym{ACRDA}{short = ACRDA ,long = asynchronous contention resolution diversity ALOHA}
\DeclareAcronym{AMP}{short = AMP ,long = approximate message passing}
\DeclareAcronym{AWGN}{short = AWGN ,long = additive white Gaussian noise}
\DeclareAcronym{CDF}{short = CDF ,long = cumulative distribution function}
\DeclareAcronym{CoSaMP}{short = CoSaMP ,long = compressive sampling matching pursuit}
\DeclareAcronym{CRA-CC}{short = CRA-CC ,long = CRA-convolutional code}
\DeclareAcronym{CRA-SH}{short = CRA-SH ,long = CRA-shannon bound}
\DeclareAcronym{CRA}{short = CRA ,long = contention resolution ALOHA}
\DeclareAcronym{CRDSA}{short = CRDSA ,long = contention resolution diversity slotted ALOHA}
\DeclareAcronym{CRDSA++}{short = CRDSA++ ,long = contention resolution diversity slotted ALOHA++}
\DeclareAcronym{CRI}{short = CRI ,long = contention resolution interval}
\DeclareAcronym{CS}{short = CS ,long = compressed sensing}
\DeclareAcronym{CSA}{short = CSA ,long = coded slotted ALOHA}
\DeclareAcronym{CSI}{short = CSI ,long = channel state information}
\DeclareAcronym{DAMA}{short = DAMA ,long = demand assigned multiple access}
\DeclareAcronym{DSA}{short = DSA ,long = diversity slotted ALOHA}
\DeclareAcronym{DSSS}{short = DSSS ,long = direct sequence spread spectrum}
\DeclareAcronym{E_SSA}{short = E-SSA ,long = enhanced spread spectrum ALOHA}
\DeclareAcronym{ECRA}{short = ECRA ,long = enhanced contention resolution ALOHA}
\DeclareAcronym{ECRA-SC}{short = ECRA-SC ,long = ECRA selection combining}
\DeclareAcronym{ECRA-MRC}{short = ECRA-MRC ,long = ECRA maximal-ratio combining}
\DeclareAcronym{EGC}{short = EGC ,long = equal-gain combining}
\DeclareAcronym{FEC}{short = FEC ,long = forward error correction}
\DeclareAcronym{gOMP}{short = gOMP ,long = generalized \ac{OMP}}
\DeclareAcronym{GEO}{short = GEO ,long = geostationary orbit}
\DeclareAcronym{GMAC}{short = GMAC ,long = Gaussian multiple access channel}
\DeclareAcronym{HAP}{short = HAP ,long = high-altitude platform}
\DeclareAcronym{IC}{short = IC ,long = interference cancellation}
\DeclareAcronym{IoT}{short = IoT ,long = Internet of things}
\DeclareAcronym{IRA}{short = IRA ,long = irregular repetition ALOHA}
\DeclareAcronym{IRCRA}{short = IRCRA ,long = irregular repetition contention resolution ALOHA}
\DeclareAcronym{IRSA}{short = IRSA ,long = irregular repetition slotted ALOHA}
\DeclareAcronym{LASSO}{short = LASSO ,long = least absolute shrinkage and selection operator}
\DeclareAcronym{LDPC}{short = LDPC ,long = low-density parity-check}
\DeclareAcronym{LEO}{short = LEO ,long = low-Earth orbit}
\DeclareAcronym{M2M}{short = M2M ,long = machine-to-machine}
\DeclareAcronym{MAC}{short = MAC ,long = medium access}
\DeclareAcronym{MAP}{short = MAP ,long = maximum-a-posteriori}
\DeclareAcronym{MF}{short = MF ,long = matched filter}
\DeclareAcronym{MF-TDMA}{short = MF-TDMA ,long = multi-frequency time division multiple access}
\DeclareAcronym{MRC}{short = MRC ,long = maximal-ratio combining}
\DeclareAcronym{MUD}{short = MUD ,long = multiuser detection}
\DeclareAcronym{NTN}{short = NTN ,long = non-terrestrial network}
\DeclareAcronym{OMP}{short = OMP ,long = orthogonal matching pursuit}
\DeclareAcronym{PDF}{short = PDF ,long = probability density function}
\DeclareAcronym{PER}{short = PER ,long = packet error rate}
\DeclareAcronym{PLR}{short = PLR ,long = packet loss rate}
\DeclareAcronym{QPSK}{short = QPSK ,long = quadrature phase-shift keying}
\DeclareAcronym{RA}{short = RA ,long = random access}
\DeclareAcronym{RCB}{short = RCB ,long = random coding bound}
\DeclareAcronym{ROC}{short = ROC ,long = receiver operating characteristic}
\DeclareAcronym{RTT}{short = RTT ,long = round trip time}
\DeclareAcronym{SA}{short = SA , long = slotted ALOHA}
\DeclareAcronym{SB}{short = SB ,long = Shannon bound}
\DeclareAcronym{SC}{short = SC ,long = selection combining}
\DeclareAcronym{SIC}{short = SIC ,long = successive interference cancellation}
\DeclareAcronym{SNIR}{short = SNIR ,long = signal-to-noise and interference ratio}
\DeclareAcronym{SINR}{short = SINR ,long = signal-to-interference and noise ratio}
\DeclareAcronym{SNR}{short = SNR ,long = signal-to-noise ratio}
\DeclareAcronym{SP}{short = SP ,long = subspace pursuit}
\DeclareAcronym{TDMA}{short = TDMA ,long = time division multiple access}
\DeclareAcronym{UCP}{short = $\Code$-UCP ,long = $\Code$-unresolvable collision pattern}
\DeclareAcronym{UMA}{short = UMA ,long = unsourced multiple access}
\DeclareAcronym{VF}{short = VF ,long = virtual frame}

\IEEEoverridecommandlockouts

\begin{document}
%
\title{An Investigation of the Compressed Sensing Phase in Unsourced Multiple Access}

\author{\IEEEauthorblockN{Federico Clazzer, Farouk Amri, Marcel Grec}
\IEEEauthorblockA{Institute of Communications and Navigation\\
German Aerospace Center (DLR)\\
Email: \{federico.clazzer, farouk.amri, marcel.grec\}@dlr.de}
\thanks{F. Clazzer, F. Amri and M. Grec acknowledge the financial support by the Federal Ministry of Education and Research of Germany in the programme of “Souverän. Digital. Vernetzt.” Joint project 6G-RIC, project identification number: 16KISK022.}}

\maketitle

\begin{abstract}
A vast population of low-cost low-power transmitters sporadically sending small amounts of data over a common wireless medium is one of the main scenarios for \ac{IoT} data communications. At the medium access, the use of grant-free solutions may be preferred to reduce overhead even at the cost of multiple-access interference. \Ac{UMA} has been recently established as relevant framework for energy efficient grant-free protocols. The use of a \ac{CS} transmission phase is key in one of the two main classes of \ac{UMA} protocols, yet little attention has been posed to sparse greedy algorithms as \ac{OMP} and its variants. We analyze their performance and provide relevant guidance on how to optimally setup the \ac{CS} phase. Minimum average transmission power and minimum number of channel uses are investigated together with the performance in terms of \ac{ROC}. Interestingly, we show how the basic \ac{OMP} and \ac{gOMP} are the most competitive algorithms in their class.
\end{abstract}

\newtheorem{definition}{Definition}
\newtheorem{prop}{Proposition}
\newtheorem{remark}{Remark}

\renewcommand{\Pr}{\ensuremath{\text{Pr}}}

\newcommand{\nbit}{b}
\newcommand{\nbitpre}{\nbit_{p}}
\newcommand{\nbitdata}{\nbit_{d}}

\newcommand{\intpre}{d_{p}}

\newcommand{\ntot}{N_t}
\newcommand{\nme}{N_p}
\newcommand{\ndata}{N_c}

\newcommand{\ptar}{p_d}
\newcommand{\pfal}{p_f}

\newcommand{\txvec}{\bm x}
\newcommand{\txenc}{\tilde{\txvec}}
\newcommand{\txvecKa}{\bar{\txvec}}
\newcommand{\txvecKaest}{\hat{\txvec}}
\newcommand{\vecKa}{\bm z}
\newcommand{\vecKaest}{\hat{\vecKa}}
\newcommand{\mxs}{\bm A}
\newcommand{\mxsest}{\bar{\bm A}}
\newcommand{\cols}{\bm a}

\newcommand{\ns}{\bm{n}}
\newcommand{\rx}{\bm{y}}

\newcommand{\res}{\bm{r}}
\newcommand{\supS}{\mathcal{T}}
\newcommand{\supSPr}{\bar \supS}
\newcommand{\cnt}{c}
\newcommand{\ind}{t}
\newcommand{\indmx}{\bar{\ind}}
\newcommand{\lcnt}{l}
\newcommand{\indgomp}{L}

\newcommand{\naus}{K_{a}}

\newcommand{\pwo}{P_{1}}
\newcommand{\pwt}{P_{2}}
\newcommand{\nspw}{N}

\newcommand{\ecs}{E_p}
\newcommand{\edata}{E_d}
\newcommand{\fren}{\alpha}
\newcommand{\symd}{T_s}


%
\IEEEpeerreviewmaketitle

\vspace{-6pt}
\section{Introduction}
\acresetall

\IEEEPARstart{T}{he} increase in \ac{IoT} traffic and services is calling for the definition of novel communication methods able to cope with the peculiarity of machine-generated data and yet, efficiently exploit the scarce resources available. In some embodiment of \ac{IoT}, low-cost low-power transmitters shall be able to sporadically transmit small amount of data over a shared wireless medium. The number of such devices is very large, but only a small fraction is active at any given time \cite{Durisi2016_IntroMMC}. This type of data traffic \--- very different from broadband user-generated data \--- requires novel approaches to communications.

At the \ac{MAC}, scheduling of traffic may become particularly inefficient as the overhead becomes comparable to the data. Grant-free solutions may be preferred, at the cost of multiple-access interference. From classic \ac{RA} solutions \cite{Abramson1970}, recently we observed a large growth in novel uncoordinated access methods. They sparkled from a recent work \cite{Polyanskiy2017} where an achievability bound on the minimum energy per bit needed to support a certain number of active users in a grant-free access has been derived and the \ac{UMA} framework has been established. Differently from previous information-theoretic works, the bound embeds the finite-length nature of messages which is particularly relevant in \ac{IoT} and, poses the focus on energy efficiency.

In the quest to find low-complexity solutions, a large body of research has recently emerged. Two main directions have been undertaken. One solution splits the user message in two parts. Both sections are separately encoded and the Kroenecker product is computed. This sequence is what is transmitted over the channel, see e.g. \cite{Pradhan2020}. A second class of solutions rely on the use of \ac{CS} encoding. Two main approaches are found in literature. In the first approach, the message is split into a number of fragments and an outer code is used for signalling which fragments belong to the same message \cite{Amalladinne2020,Fengler2020}. Then those fragments are transmitted through the channel by encoding them with \ac{CS}. In the second approach, the message is split in only two fragments of possibly different sizes. The first part is transmitted via \ac{CS} and it is also used to signal some side-information exploited in the transmission of the second fragment \cite{Pradhan2022,Truhcachev2020,Zheng2020}, see e.g. Fig.~\ref{fig:CS_UMA} for the transmitter architecture. For \ac{CS} decoding, the proposed schemes adopt either basis pursuit or \ac{LASSO} \cite{Chen1994_BP,Tibshirani1996_LASSO} or \ac{AMP} \cite{Donoho2010_AMP}.

In the \ac{UMA} setting, little attention has been given to the use of sparse greedy recovery algorithms as, e.g., \ac{OMP} \cite{Pati1993_OMP}. These \ac{CS} algorithms have the advantage of low complexity and relatively good recovery performance. Hence, in this work we will investigate their use for \ac{UMA} schemes. After a brief review of the main variants of \ac{OMP}, we will analyze their performance in terms of relevant performance metrics. In particular, setting a target detection probability to $\ptar=99.9\%$ we will evaluate the minimum average transmission power for the \ac{CS} phase, or the minimum number of channel uses, both as a function of the number of active users.

\section{System Model}

\begin{figure}
\centering
\includegraphics[width=0.55\columnwidth]{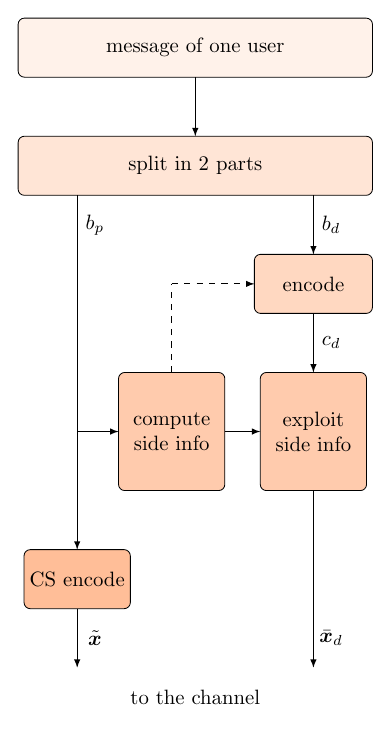}
\caption{Considered transmitter architecture. The message of every user is split in two parts. The first part entails the first $\nbitpre$ bits of the message and are encoded with \ac{CS} while, the second part is composed by remaining $\nbitdata$ bits. The $\nbitdata$ bits are also encoded and side information is embedded before transmission. The side information can be in the form of a user-specific spreading sequence as in \cite{Truhcachev2020, Zheng2020} or a user-specific permutation as in \cite{Pradhan2022} and, it is indicated by the first $\nbitpre$ bits.}
\label{fig:CS_UMA}
\end{figure}

Our focus is on an \ac{UMA} system where $\naus$ active users are interested in transmitting $\nbit$ bits each, over the \ac{GMAC}~\cite{Polyanskiy2017} without coordination. As depicted in Fig.~\ref{fig:CS_UMA} and, similarly to \cite{Truhcachev2020, Zheng2020, Pradhan2022}, we assume that the message is split in two parts. The first part entails $\nbitpre$ bits of the message while, the second part is composed by remaining $\nbitdata$ bits, so that $\nbit= \nbitpre + \nbitdata$. In the aforementioned works,  $\nbitpre$ bits are both transmitted with \ac{CS} and are also used to compute some side information exploited in the transmission of the second $\nbitdata$ bits. For example, in \cite{Truhcachev2020} side information is a spreading sequence and a permutation, while encoding is done by a repetition code. In \cite{Zheng2020}, the side information entails code length, power level and an interleaver sequence. Note that in this case, side information is exploited also at the channel encoder. Encoding is done by exploiting polar codes. In \cite{Pradhan2022} side information includes the number of repetitions and a permutation, while encoding is done with an \ac{LDPC}. 

In this contribution, our focus is on the \ac{CS} transmission, hence, we will investigate the performance of different \ac{CS} decoding algorithms as well as optimize their configuration for the \ac{UMA}. Since we are exclusively interested in the \ac{CS} part of the protocol, we can concentrate on the transmission of the first $\nbitpre$ bits only.

\subsection{Compressed Sensing Encoding}

The encoding of the first $\nbitpre$ bits follows three steps:
\begin{enumerate}
\item the $\nbitpre$ bits are converted in an integer $\intpre \in \left[0,2^{\nbitpre}-1\right]$.
\item a binary vector $\txvec \in \mathbb{B}^{2^{\nbitpre} \times 1}$ whose entries are all zeros apart in position $\intpre$ is built, i.e. ${\txvec = \left[\underset{0}{0},\dots,\underset{\intpre-1}{0},\underset{\intpre}{1},\underset{\intpre+1}{0},\dots,\underset{2^{\nbitpre}-1}{0}\right]}$.\vspace{5pt}
\item the binary vector $\txvec$ multiplies the sensing matrix ${\mxs \in \mathbb{R}^{\nme \times 2^{\nbitpre}}}$ to generate the transmission vector ${\txenc \in \mathbb{R}^{\nme \times 1}}$, as $\txenc = \mxs \txvec$.
\end{enumerate}

\begin{remark}
Note that the sensing matrix $\mxs$ is common to all transmitters and it is also known by the receiver. Additionally, we assume that the columns of $\mxs$ are normalized to $\nme \pwo$, i.e. $\|\cols_j\|^2=\nme \pwo$ with $j \in \left[0,\dots,2^{\nbitpre}-1\right]$. $\pwo$ is the average power for the transmission of the $\nme$ symbols corresponding to the $\nbitpre$ bits.
\end{remark}

\subsection{Channel and \ac{CS} Decoding}

The $\naus$ users concurrently transmit over the same bandwidth in a \ac{GMAC}. Hence the receiver observes
\begin{equation}
\label{eq:rxsig}
\rx = \sum_{i=1}^{\naus} \txenc_i + \ns = \mxs \sum_{i=1}^{\naus} \txvec_i + \ns = \mxs \txvecKa + \ns
\end{equation}
with $\rx \in \mathbb{R}^{\nme \times 1}$, $\txvecKa = \sum_{i=1}^{\naus} \txvec_i \in \mathbb{B}^{2^{\nbitpre} \times 1}$ a $\naus$-sparse binary vector\footnote{We disregard the probability that two users have in common the first $\nbitpre$ bits of the message, which would result in the reduction of the sparsity. For realistic dimension of at least $10$ bits, this probability is below $10^{-3}$ assuming equi-probable messages.} and $\ns \sim \mathcal{N}\left(0,1\right)$ the noise vector.

The task of the receiver is to estimate $\txvecKa$, or better its support, having access to the received signal vector $\rx$ and exploiting the side information of the sensing matrix $\mxs$. Eq.~\eqref{eq:rxsig} is an instance of the noisy \acl{CS} problem, sometimes also referred to as to robust \acl{CS} \cite{Candes2008Cs}. We shall observe that in absence of any additional information on $\txvecKa$, eq.~\eqref{eq:rxsig} is well-posed only if $\nme \geq 2^{\nbitpre}$. However, since $\txvecKa$ is sparse there are algorithms able to provide an accurate estimate of $\txvecKa$ even when $\nme \ll 2^{\nbitpre}$.

There are four main classes of \emph{sparse recovery algorithms}~\cite{Choi2017_CSTut} that aim at estimating the support of $\txvecKa$:
\begin{itemize}
\item \emph{convex optimization}, where the $l_1$-norm minimization is used as proxy for the $l_0$-norm minimization to reduce the computational complexity. In this class, basis pursuit~\cite{Chen1994_BP} and \ac{LASSO}~\cite{Tibshirani1996_LASSO} are the most well-known algorithms.
\item \emph{Greedy algorithms} successively identify the subset of the support. To this class belong \ac{OMP}~\cite{Pati1993_OMP}, \ac{gOMP}~\cite{Wang2012_GOMP}, \ac{CoSaMP}~\cite{Needell2009_CoSaMP} and \ac{SP}~\cite{Dai2009_SP} for example. In-depth discussion on the aforementioned algorithms will follow as they are the focus of this investigation.
\item \emph{Iterative algorithms} apply a non-linear operator that will set to zero all but a pre-defined number of elements in the estimated sparse vector. This operation is applied in an iterative fashion and, convergence can be guaranteed asymptotically, if the sensing matrix undergoes some properties~\cite{Blumensath2009_IHT}. Belonging to this class, we can also mention \ac{AMP}~\cite{Donoho2010_AMP}.
\item \emph{Statistical sparse recovery} assumes the sparse vector $\txvecKa$ to be a random vector. Bayesian inference is then leveraged to compute a suitable estimate of the support of $\txvecKa$. In this class we can mention \ac{MAP} estimation with Laplacian prior~\cite{Figueiredo2001_MAP}, sparse Bayesian learning~\cite{Wipf2004_SBL} and Bayesian compressed sensing~\cite{Ji2008_BCS}.
\end{itemize}

Our focus is on the investigation of greedy algorithms, due to their limited complexity with respect to the other sparse recovery algorithms classes. Before investigating their performance in the \ac{UMA} setting, in the next section we will review how they recover an estimate of the support of $\txvecKa$.

\section{Sparse Greedy Recovery Algorithms for \ac{UMA}}
\label{sec:Greedysparse}

In this section we review the steps followed by four greedy sparse recovery algorithms, namely \ac{OMP}, \ac{gOMP}, \ac{CoSaMP} and \ac{SP} to estimate the support of $\txvecKa$.

\begin{algorithm}
\caption{\ac{OMP}}\label{alg:OMP}
\hspace*{\algorithmicindent} \textbf{Input:} $\rx$, $\mxs$, $\naus$\\
\hspace*{\algorithmicindent} \textbf{Output:} $\txvecKaest$ \Comment{estimate of $\txvecKa$}
\begin{algorithmic}[1]
\Procedure{}{}
\State $\res_0 = \rx$ \Comment{initialize the residual}
\State $\supS_0 \gets \emptyset$ \Comment{initialize the support set}
\State $\mxsest_0 \gets []$ \Comment{initialize the estimated matrix}
\State $\txvecKaest_0 \gets \bm{0}$ \Comment{initialize the estimated support}
\State $\cnt = 1$ \Comment{initialize the iterations counter}
\While{$\cnt < \left(\naus + 1\right)$}
    \State $\ind_\cnt = \underset{j=0,\dots,2^{\nbitpre}-1}{\arg \max} |\langle \res_{\cnt-1},\cols_j  \rangle|$ \Comment{corr.}
    \State $\supS_{\cnt} = \supS_{\cnt-1} \cup \{\ind_\cnt\}$ \Comment{update index set}
    \State $\vecKaest_{\cnt} = \arg \min_{\vecKa}\|\rx-\mxsest_{\supS_{\cnt}} \vecKa\|$ \Comment{least squares}
    \State $\res_{\cnt} = \rx - \mxsest_{\supS_{\cnt}} \vecKaest_{\cnt}$ \Comment{update residual}
    \State $\txvecKaest_{\cnt}(\ind_\cnt) \gets 1$ \Comment{update est. support}
    \State $\cnt = \cnt + 1$
\EndWhile
\EndProcedure
\end{algorithmic}
\end{algorithm}

In Algorithm~\ref{alg:OMP} \ac{OMP} is presented. The most important step is to find the maximum correlation between the residual and the columns of the sensing matrix $\mxs$. The index set of the selected columns is then augmented. By computing least squares, $\vecKaest_{\cnt}$ the approximation of $\txvecKa$ can be evaluated, which is then exploited to update the residual. Before iterating, the selected support index is stored in the estimated vector $\txvecKaest$.
\begin{algorithm}
\caption{\ac{gOMP}}\label{alg:gOMP}
\hspace*{\algorithmicindent} \textbf{Input:} $\rx$, $\mxs$, $\naus$, $\indgomp$\\
\hspace*{\algorithmicindent} \textbf{Output:} $\txvecKaest$ \Comment{estimate of $\txvecKa$}
\begin{algorithmic}[1]
\Procedure{}{}
\State $\res_0 = \rx$ \Comment{initialize the residual}
\State $\supS_0 \gets \emptyset$ \Comment{initialize the support set}
\State $\mxsest_0 \gets []$ \Comment{initialize the estimated matrix}
\State $\txvecKaest \gets \bm{0}$ \Comment{initialize the estimated support}
\State $\cnt = 1$ \Comment{initialize the iterations counter}
\While{$\cnt < \left(\naus + 1\right)$ and $\|\res_{\cnt}\|>\epsilon$}
    \State $\lcnt = 0$
    \While{$\lcnt < \indgomp$}
        \State $\ind_\lcnt = \underset{j=0,\dots,2^{\nbitpre}-1 \backslash\left\{\ind_0,\dots,\ind_{\lcnt-1}\right\}}{\arg \max} |\langle \res_{\cnt-1},\cols_j  \rangle|$
        \State $\supS_{\cnt} = \supS_{\cnt-1} \cup \{\ind_\lcnt\}$ \Comment{update index set}
        \State $\lcnt = \lcnt + 1$
    \EndWhile
    \State $\vecKaest_{\cnt} = \arg \min_{\vecKa}\|\rx-\mxsest_{\supS_{\cnt}} \vecKa\|$ \Comment{least squares}
    \State $\res_{\cnt} = \rx - \mxsest_{\supS_{\cnt}} \vecKaest_{\cnt}$ \Comment{update residual}

    \State $\cnt = \cnt + 1$
\EndWhile
\State $\left\{\indmx_1,\dots,\indmx_{\naus}\right\} \gets |\vecKaest_{\cnt|\naus}|$ \Comment{sel. $\naus$ largest ind.}
\State $\supSPr \gets \left\{\indmx_1,\dots,\indmx_{\naus}\right\}$
\State $\txvecKaest(\supSPr) \gets 1$ \Comment{estimated support}
\EndProcedure
\end{algorithmic}
\end{algorithm}

By observing that \ac{OMP} will require a large number of iterations as the sparsity increases, \ac{gOMP} \--- see Algorithm~\ref{alg:gOMP} \--- has been proposed as possible solution to reduce the number of iterations. By selecting an integer $\indgomp$ that represents how many columns can be selected at each iteration, \ac{gOMP} should be able to converge faster than \ac{OMP}. In fact, if more than one column belong to the support of $\txvecKa$, one or more iterations can be spared. Note that, even if the columns of $\mxs$ selected exceed the sparsity level, they are anyhow kept for the iterative update of the residual.
\begin{algorithm}
\caption{\ac{CoSaMP}}\label{alg:cosamp}
\hspace*{\algorithmicindent} \textbf{Input:} $\rx$, $\mxs$, $\naus$\\
\hspace*{\algorithmicindent} \textbf{Output:} $\txvecKaest$ \Comment{estimate of $\txvecKa$}
\begin{algorithmic}[1]
\Procedure{}{}
\State $\res_0 = \rx$ \Comment{initialize the residual}
\State $\supS_0 \gets \emptyset$ \Comment{initialize the support set}
\State $\mxsest_0 \gets []$ \Comment{initialize the estimated matrix}
\State $\txvecKaest \gets \bm{0}$ \Comment{initialize the estimated support}
\State $\cnt = 1$ \Comment{initialize the iterations counter}
\While{$\cnt < \left(\naus + 1\right)$ and $\|\res_{\cnt}\|>\epsilon$}
    \State $\lcnt = 0$
    \While{$\lcnt < 2 \naus$}
        \State $\ind_\lcnt = \underset{j=0,\dots,2^{\nbitpre}-1 \backslash\left\{\ind_0,\dots,\ind_{\lcnt-1}\right\}}{\arg \max} |\langle \res_{\cnt-1},\cols_j  \rangle|$
        \State $\supS_{\cnt} = \supS_{\cnt-1} \cup \{\ind_\lcnt\}$ \Comment{update index set}
        \State $\lcnt = \lcnt + 1$
    \EndWhile
    \State $\vecKaest_{\cnt} = \arg \min_{\vecKa}\|\rx-\mxsest_{\supS_{\cnt}} \vecKa\|$ \Comment{least squares}
    \State $\left\{\indmx_1,\dots,\indmx_{\naus}\right\} \gets |\vecKaest_{\cnt|\naus}|$ \Comment{sel. $\naus$ largest ind.}
    \State $\supSPr \gets \left\{\indmx_1,\dots,\indmx_{\naus}\right\}$

    \State $\res_{\cnt} = \rx - \mxsest_{\supSPr} \vecKaest_{\supSPr}$ \Comment{update residual}
    \State $\supS_{\cnt} = \supSPr$
    \State $\cnt = \cnt + 1$
\EndWhile
\State $\txvecKaest(\supSPr) \gets 1$ \Comment{estimated support}
\EndProcedure
\end{algorithmic}
\end{algorithm}

\ac{CoSaMP} and \ac{SP} take a step even further and select $2\naus$ and $\naus$ column at each iteration respectively. These columns together with the support set at the previous iteration, whose size is $\naus$ in both cases, are utilized to compute the least squares approximation of $\txvecKa$. Before the residual is computed, the indexes corresponding to the largest $\naus$ columns are kept, while the remaining are discarded. Hence, the support set at the end of each iteration has the size of the sparsity $\naus$. We summarize \ac{CoSaMP} and \ac{SP} in Algorithms~\ref{alg:cosamp} and \ref{alg:sp} respectively.
\begin{algorithm}
\caption{\ac{SP}}\label{alg:sp}
\hspace*{\algorithmicindent} \textbf{Input:} $\rx$, $\mxs$, $\naus$\\
\hspace*{\algorithmicindent} \textbf{Output:} $\txvecKaest$ \Comment{estimate of $\txvecKa$}
\begin{algorithmic}[1]
\Procedure{}{}
\State $\res_0 = \rx$ \Comment{initialize the residual}
\State $\supS_0 \gets \emptyset$ \Comment{initialize the support set}
\State $\mxsest_0 \gets []$ \Comment{initialize the estimated matrix}
\State $\txvecKaest \gets \bm{0}$ \Comment{initialize the estimated support}
\State $\cnt = 1$ \Comment{initialize the iterations counter}
\While{$\cnt < \left(\naus + 1\right)$ and $\|\res_{\cnt}\|>\epsilon$}
    \State $\lcnt = 0$
    \While{$\lcnt < \naus$}
        \State $\ind_\lcnt = \underset{j=0,\dots,2^{\nbitpre}-1 \backslash\left\{\ind_0,\dots,\ind_{\lcnt-1}\right\}}{\arg \max} |\langle \res_{\cnt-1},\cols_j  \rangle|$
        \State $\supS_{\cnt} = \supS_{\cnt-1} \cup \{\ind_\lcnt\}$ \Comment{update index set}
        \State $\lcnt = \lcnt + 1$
    \EndWhile
    \State $\vecKaest_{\cnt} = \arg \min_{\vecKa}\|\rx-\mxsest_{\supS_{\cnt}} \vecKa\|$ \Comment{least squares}
    \State $\left\{\indmx_1,\dots,\indmx_{\naus}\right\} \gets |\vecKaest_{\cnt|\naus}|$ \Comment{sel. $\naus$ largest ind.}
    \State $\supSPr \gets \left\{\indmx_1,\dots,\indmx_{\naus}\right\}$

    \State $\res_{\cnt} = \rx - \mxsest_{\supSPr} \vecKaest_{\supSPr}$ \Comment{update residual}
    \State $\supS_{\cnt} = \supSPr$
    \State $\cnt = \cnt + 1$
\EndWhile
\State $\txvecKaest(\supSPr) \gets 1$ \Comment{estimated support}
\EndProcedure
\end{algorithmic}
\end{algorithm}

In the next Section we present numerical results for the four sparse greedy recovering algorithms in the \ac{UMA} setting.

\section{Numerical Results}

Before entering in the numerical investigation, we list the assumptions and the setting considered. As per the definition of the \ac{UMA} in \cite{Polyanskiy2017}, we assume that each user needs to transmit $\nbit=100$~bits which are mapped to a total of ${\ntot=30\,000}$ channel uses. Similarly to \cite{Pradhan2022}, we split the message in two parts, the first is composed by $\nbitpre=15$~bits, while the second entails the remaining $\nbitdata=85$~bits. $\nme$ channel uses are allotted for the first part of the message, while the remaining $\ndata=\ntot-\nme$ are reserved for the transmission of the second $\nbitdata$~bits. Recall that the first $\nbitpre$~bits are encoded via \ac{CS} and are transmitted over the \ac{GMAC} as per eq.~\eqref{eq:rxsig}. The sensing matrix $\mxs$ is constructed from a square discrete cosine transform matrix $\mathbb{R}^{2^{\nbitpre} \times 2^{\nbitpre}}$, by selecting uniformly at random $\nme$ rows. The average received power per symbol in the \ac{CS} phase is $\pwo$. It also corresponds to the average per-symbol \ac{SNR}, as the noise power is normalized to one, i.e. $\nspw=1$ so $\pwo=\pwo/ \nspw$. We further assume that the average per-symbol \ac{SNR} for the second part of the message is $\pwt$. At the receiver, we will decode the concurrent transmissions by employing one of the four greedy sparse recovery algorithms presented in Sec.~\ref{sec:Greedysparse}. We aim at a recovery probability of $\ptar=99.9\%$, i.e. a maximum of one in thousand messages can be wrongly detected after \ac{CS} decoding. The corresponding error probability for the \ac{CS} part is thus $0.1\%$, well below the per user error probability of $5\%$ of the \ac{UMA} schemes (see e.g. \cite{Polyanskiy2017, Pradhan2022}). In this way, we are ensuring that the \ac{CS} decoding will not be the bottleneck of the \ac{UMA} protocol.

We denote with $\ecs$ the energy dedicated to the transmission of the first part of the message, and with $\edata$ the energy dedicated to the transmission of the second part of the message. The duration of a symbol is $\symd$. We define $\fren$ as the ratio between $\ecs$ and $\edata$,

\begin{equation}
\label{eq:alpha}
\fren \triangleq \frac{\ecs}{\edata} = \frac{\pwo \symd \nme}{\pwt \symd \ndata}.
\end{equation}

Similarly to \cite{Pradhan2022} it holds
\begin{equation}
\label{eq:Eb_N0}
\frac{E_b}{N_0} = \frac{\pwo \nme + \pwt \ndata}{2 \nbit} = \pwo \frac{\fren \nme + \nme}{2 \fren \nbit }.
\end{equation}
where $\frac{E_b}{N_0}$ is the energy per bit.

\begin{remark}
    When $\fren=1$, the energy dedicated to the \ac{CS} transmission is the same of the energy reserved to the second part of the message, i.e. $\ecs=\edata$. In this case, eq.~\eqref{eq:Eb_N0} simplifies to $\frac{E_b}{N_0} = \pwo \frac{\nme}{\nbit}$.
\end{remark}

\subsection{Minimum power $\pwo$}

In the first results, we fix the number of channel uses for the \ac{CS} part to $\nme = 2\,000$ as per \cite{Pradhan2022}. We are interested in evaluating the minimum average power per transmitted symbol in the \ac{CS} phase $\pwo$, that achieves the target recovery probability $\ptar=99.9\%$.

\begin{figure}
\centering
\includegraphics[width=.9\columnwidth]{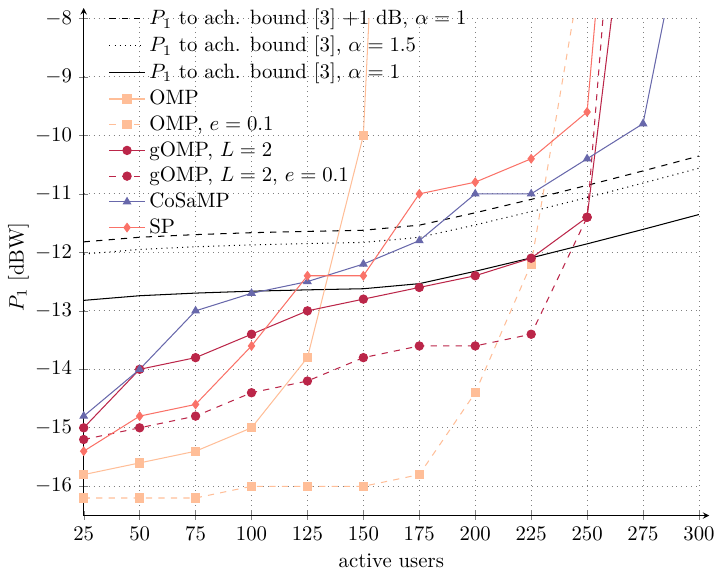}
\caption{Active users vs. minimum $\pwo$ that guarantees the target recovery probability $\ptar=99.9\%$. We compare the four greedy sparse recovery algorithms \ac{OMP}, \ac{gOMP}, \ac{CoSaMP} and \ac{SP}. For \ac{OMP} and \ac{gOMP} we also provide to the \ac{CS} algorithm a sparsity level in excess by $10\%$ (dashed lines). The black solid line represents the maximum $\pwo$ that needs to be achieved by the \ac{CS} phase so that the overall per-user error probability does not exceed $5\%$ and assuming $\fren=1$. The black dashed line is the same but with a $1$~dB backoff. The black dotted line instead assumes $\fren=1.5$ and no backoff.}
\label{fig:UMA_CS_SNR}
\end{figure}

The results are collected in Fig.~\ref{fig:UMA_CS_SNR}. On the x-axis we depict the number of active users, $\naus$, while on the y-axis the minimum $\pwo$ that guarantees the target recovery probability is found. Note that for \ac{gOMP} we present results only for $\indgomp=2$ as we empirically observed that this is the best choice for $\indgomp$. Up to $\naus=125$, the basic \ac{OMP} provides the best performance with the smallest $\pwo$ among all four \ac{CS} algorithms. In particular, it achieves $\pwo \approx -16$~dBW for $\naus=25$. Beyond this sparsity, \ac{gOMP} achieves the best performance up to $250$ active users when \ac{CoSaMP} takes over as the best decoding algorithm.
In order to provide a reference to the \ac{UMA} setting, we depicted some additional curves in Fig.~\ref{fig:UMA_CS_SNR}. The black solid line represents the maximum $\pwo$ that needs to be achieved by the \ac{CS} phase so that the overall per-user error probability does not exceed $5\%$ and assuming $\fren=1$, i.e. $\ecs = \edata$. In other words, assume that you designed a very good \ac{UMA} scheme able to match the achievability bound of \cite{Polyanskiy2017}, and you allow the same energy to be spent in the \ac{CS} and in the second part of the transmission then, the maximum $\pwo$ is represented by the dashed line. Whenever the \ac{CS} decoding algorithm reliably works\footnote{Here we assume that $\ptar=99.9\%$ suffices to provide sufficient reliability to the \ac{CS} phase.} with a $\pwo$ smaller than the solid line, than it can operate in conjunction with a scheme able to achieve the bound. We can observe that while \ac{SP} and \ac{CoSaMP} can be adopted up to $\naus=100$, \ac{OMP} can be used up to $\naus=125$ and \ac{gOMP} even up to $\naus=225$ users. If the sparsity range guaranteed by the \ac{CS} algorithm is not sufficient, some options remain. One can increase the energy reserved to the \ac{CS} phase, see dotted line. Or assume to pay a backoff w.r.t. the achievability bound due to imperfections in the scheme. The black dashed line for example, assumes $\fren=1$ but a $1$~dB backoff w.r.t. the achievability bound. In both cases, the sparsity range is now expanded to $\naus=150$ for \ac{SP}, $\naus=175$ for \ac{CoSaMP} and $\naus=250$ for \ac{gOMP}. Finally, one could also increase the sparsity level passed to the \ac{CS} algorithm. In this way, a higher false alarm probability is permitted. The wrongly selected messages can be discarded in the second phase when the second part of the message is processed. In fact, error detection can be exploited to discard the messages in excess. Accounting for a sparsity in excess of 10\%, \ac{OMP} becomes the best performing scheme and can support a sparsity up to $\naus=225$ (see dashed curve denoted with \ac{OMP}, $e=0.1$). Similarly, also \ac{gOMP} benefits for a sparsity excess of 10\%. Going beyond this level, does not bring particular advantages we observed empirically. From this investigation, it appears that $\nme = 2\,000$ is a reasonable choice for a sparsity level of up to $\naus=150$ ($\naus=125$ if one uses \ac{OMP}). Nonetheless, one could ask the relevant question of whether for low sparsity levels, the number of channel uses dedicated to \ac{CS} can be reduced.
\vspace{-2pt}
\subsection{Minimum number of channel uses $\nme$}

\begin{figure}
\centering
\includegraphics[width=.9\columnwidth]{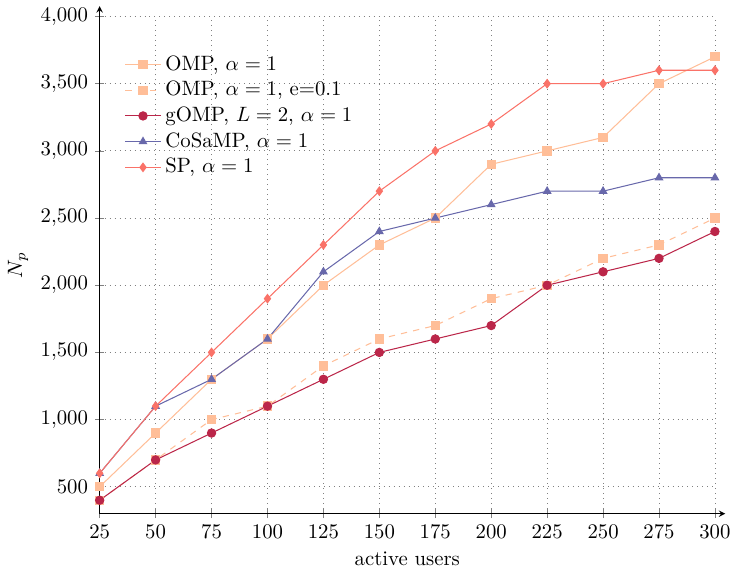}
\caption{Active users vs. minimum $\nme$ that guarantees the target recovery probability $\ptar=99.9\%$, under the assumption of equal energy, i.e. $\fren=1$ and $\frac{E_b}{N_0}$ from the achievability bound in \cite{Polyanskiy2017}. We compare the four greedy sparse recovery algorithms \ac{OMP}, \ac{gOMP}, \ac{CoSaMP} and \ac{SP}. For \ac{OMP} also provide to the \ac{CS} algorithm a sparsity level in excess by $10\%$ (dashed line).}
\label{fig:UMA_CS_sparsity}
\end{figure}

In Fig.~\ref{fig:UMA_CS_sparsity} we depict the number of channel uses dedicated to the \ac{CS} part as a function of the number of active users for the four sparse greedy algorithms, by assuming that the overall \ac{UMA} scheme is able to reliably work on the achievability bound of \cite{Polyanskiy2017} and targets $\ptar=99.9\%$. In particular, from eq.~\eqref{eq:Eb_N0}, we fix $E_b/N_0$ from the achievability bound in \cite{Polyanskiy2017}, we set $\fren=1$ and for each $\nme$ we compute the corresponding $\pwo$ at which the \ac{CS} algorithm shall operate. The minimum $\nme$ for which $\ptar=99.9\%$ is the one depicted in the Figure.

Differently from the previous results, \ac{gOMP} achieves the best performance for all active users considered in the evaluation. Additionally, as we would expect from Fig.~\ref{fig:UMA_CS_SNR}, ${\nme=2\,000}$ is achieved for $\naus=225$ as this is maximum number of active users for which \ac{gOMP} lies below the achievability bound. If \ac{gOMP} is selected as \ac{CS} decoding algorithm, one could half the channel uses dedicated to the first part of the message if the maximum number of users to be supported does not exceed $100$. \ac{OMP} is the second best performing \ac{CS} algorithm, even though the benefit w.r.t. to \ac{CoSaMP} or \ac{SP} is in this case rather limited. When an excess of $10\%$ in the sparsity level is given as input to \ac{OMP} (dashed line) the performance becomes comparable with \ac{gOMP}. Instead, for \ac{gOMP} we observed that no advantage can be harvested if an excess of $10\%$ in the sparsity level is given as input. Additionally, we also observed that the required minimum number of channel uses is robust in terms of variations of $\fren$. Until now we have assumed that $\naus$ is known by the receiver and can be used as input to the sparse greedy algorithms. In reality, this value is hardly known as these schemes address grant-free communications. In the next results we modify the algorithms so that the sparsity level $\naus$ is provided as output together with the estimated support.

\subsection{\Ac{ROC}}

\begin{figure}
\centering
\includegraphics[width=.85\columnwidth]{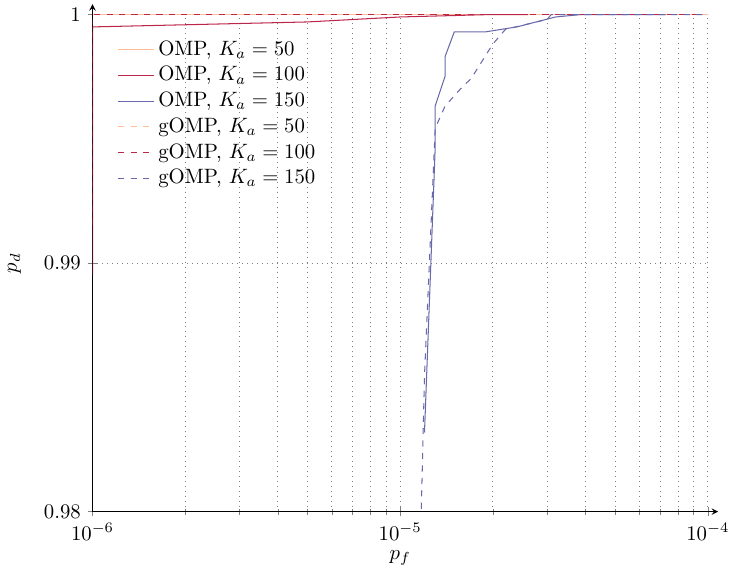}
\vspace{-1em}
\caption{\Acf{ROC} for \ac{OMP} and \ac{gOMP} and with $\naus \in \left\{50,100,150\right\}$. $\ptar$ is the detection probability while $\pfal$ is the probability that any of the $\nme-\naus$ columns of $\mxs$ is marked as active even if it was not selected by any user. We fix $\nme=2000$ channel uses, $\alpha=1$ and, $\pwo = \frac{E_b}{N_0}\frac{\nbit}{\nme}$ (see eq.~\eqref{eq:Eb_N0}), where $\frac{E_b}{N_0}$ is computed according to \cite{Polyanskiy2017} for the corresponding number of active users.}
\vspace{-1em}
\label{fig:UMA_CS_ROC}
\end{figure}

In Fig.~\ref{fig:UMA_CS_ROC} we depict the \ac{ROC}. The \ac{ROC} shows the detection probability $\ptar$ as a function of the false alarm probability $\pfal$. In our scenario, $\ptar$ represents the probability that any of the messages transmitted via \ac{CS} is correctly detected by the receiver algorithm. Conversely, $\pfal$ is the probability that one message is marked as transmitted even if it was not. In other words, $\pfal$ is the probability that any of the $\nme-\naus$ columns of $\mxs$ is marked as active even if it was not selected by any user. In the simulations, we fix $\nme=2000$ channel uses, $\alpha=1$. Additionally, $\pwo = \frac{E_b}{N_0}\frac{\nbit}{\nme}$, where $\frac{E_b}{N_0}$ is computed according to \cite{Polyanskiy2017} for the corresponding number of active users. We depict the performance for $\naus \in \left\{50,100,150\right\}$. Both \ac{OMP} and \ac{gOMP} provide a very good performance. The detection probability is $\gg 0.99$ with a false alarm rate below $10^{-6}$ for $\naus=50,100$. Increasing the number of users to $\naus=150$, the false alarm rate exceeds $10^{-5}$ for both \ac{OMP} and \ac{gOMP}. This would result in one false alarm every $\sim 8\,000$ messages. From these results, it is clear that sizing the \ac{CS} phase with $\nme=2000$ channel uses, and for $\alpha=1$, will provide enough robustness.
\section{Conclusion}
\vspace{-2pt}
In this paper we investigated the use of sparse greedy recovery algorithms in the \ac{UMA} setting. By taking inspiration from the scheme \cite{Pradhan2022}, we analyzed the performance of \ac{OMP} and some of its main variants as \ac{gOMP}, \ac{CoSaMP} and \ac{SP}. The minimum average signal power required by the \ac{CS} recovery algorithm to achieve a target detection probability of $\ptar=99.9\%$ has been studied first. We observed that \ac{OMP} and \ac{gOMP} are the best performing algorithms, providing up to a $3$~dB margin with respect to an \ac{UMA} \ac{CS} algorithm on the achievability bound of \cite{Polyanskiy2017}. This margin can be exploited in two ways: one can dedicate a higher power to the second phase of the \ac{UMA} scheme and thus help the second phase decoder, or one can reduce the number of channel uses dedicated to the \ac{CS} phase. If the latter is chosen, we showed that the number of channel uses can be lowered to $\nme=1\,000$ if the number of active users does not exceed $100$ and \ac{OMP} or \ac{gOMP} are adopted. Finally, we removed the assumption of perfect knowledge of the number of active users and provided \ac{ROC} curves for \ac{OMP} and \ac{gOMP}. Targeting a detection probability of $0.999$, we observed that the false alarm probability remains well under control for various sparsity levels. It does not exceed $2 \cdot 10^{-5}$ when we dedicate $\nme=2\,000$ channel uses, showing that both \ac{OMP} and \ac{gOMP} are very robust.
%
%
%
\bibliographystyle{IEEEtran}
\vspace{-.5ex}
\bibliography{IEEEabrv,References}

\end{document}